\documentclass[a4paper]{iopart}
\usepackage{epsf}
\newcommand{\dir}[1]{\left| #1 \right\rangle}
\newcommand{\dirc}[1]{\left\langle #1 \right|}

\newcommand{\mdir}[1]{\left\langle #1\right\rangle}
\newcommand{\adir}[2]{\left\langle #1\left|#2\right|#1\right\rangle}

\begin{document}

\title{Probabilistic aspects of Wigner function}

\author{Constantin V. Usenko}
\date{}
\address{ National Shevchenko University of Kyiv, Ukraine}
\mailto{usenko@univ.kiev.ua}

\begin{abstract}
The Wigner function of quantum systems is an effective instrument to construct the approximate classical description of the systems for which the classical approximation is possible. During the last time, the Wigner function formalism is applied as well to seek indications of specific quantum properties of quantum systems leading to impossibility of the classical approximation construction. Most of all, as such an indication the existence of negative values in Wigner function for specific states of the quantum system being studied is used. The existence of such values itself prejudices the probabilistic interpretation of the Wigner function, though for an arbitrary observable depending jointly on the coordinate and the momentum of the quantum system just the Wigner function gives an effective instrument to calculate the average value and the other statistical characteristics. 

In this paper probabilistic interpretation of the Wigner function based on coordination of theoretical-probabilistic definition of the probability density, with restrictions to a physically small domain of phase space due to the uncertainty principle, is proposed. 

\end{abstract}
\maketitle

\section{Introduction}
 One of the problems in the physical understanding of quantum phenomena
 is in probabilistic interpretation of the Wigner function. This function is one of a series of the density operator representation and has direct relation to quasiclassical interpretation of quantum phenomena. 

Interest to the Wigner function is recently renewed for quantum optic applications due to its  measurement through homodine tomography \cite{homo,homod,ManMT,Kim,homodyn,homodyn1,exphomodyn,exphomodyn2}.

Main interpretation of the Wigner function follows from density operator interpretation -  Wigner function is the tool for description of quantum states and the method for calculation of observable values.

Since probabilistic interpretation of quantum phenomena is similar to the one of the classical statistical mechanics, one might expect similar interpretation of the main tools, especially of the Wigner function, which depends upon coordinate and momentum like the phase space probability distribution in classical statistical mechanics.

It is well known that for specific states the Wigner function can take negative 
values. This is in contradiction with its possible interpretation as the 
probability density distribution. Typical understanding of this property of Wigner function is that this function gives quasiprobability only \cite{kl,Class,zurek}. Nevertheless, recently the new application of the negativity of Wigner function has arosen - existence of negative values of the Wigner function is interpreted as criterion of nonclassicality of quantum state \cite{semvog,Wexp}. Such interpretation requires the reconsideration of interrelation between the nature of quantum phenomena and the probability distribution for measured values.

Subject of my talk is the influence of the uncertainty of measured values on the probabilistic interpretation of quantum phenomena.
It is well known \cite{muga} that the Wigner function is closest to the classical probability when
average local values and local variances of a quantum observable are numerically compared
to their classical analogue.
Non-negative Wigner-type
distributions for all quantum states can be obtained \cite{cartw} by smoothing with a Gaussian
for which variance is greater than or equal to that of the minimum uncertainty, or by
integrating the Wigner function over phase space regions of the order $\hbar$. 

These peculiarities of Wigner function as of other phase space representations of the density operator wait for the probability interpretation up till now. Recent discussion \cite{drago, dragob} is
confined to statement about restriction of the accuracy of measurement only, without account of the fundamental significance of uncertainty principle for quantum phenomena. 

Here the redefinition of the probability distribution on phase space is studied in close correspondence with the uncertainty principle. It is stated that probability interpretation of the Wigner function is obtained by rejecting the Hausdorff topology of phase space. The existence of the smallest size of topology base domains eliminates the need in probability interpretation of negative value.

\section{ Density operator and Wigner function}
 
 Main tool for quantum phenomena description is the density operator. It can be a projector to state subspace for pure states, or a weighted sum of projectors for mixed states, or it has an integral representation only
\[
	\hat{\rho}=\dir{\psi}\otimes \dirc{\psi};\ \hat{\rho}=\sum{\rho_n\dir{\psi_n}\otimes \dirc{\psi_n}}; \ \hat{\rho}=\int{\rho\left(\alpha,\alpha'\right) \dir{\psi\left(\alpha\right)}\otimes \dirc{\psi\left(\alpha'\right)}d\alpha d\alpha'}.
\]
It defines the state of the studied object and produces the rule (\ref{eq4}) for calculation of the average value of a quantum observable $\hat{A}$. 

\begin{equation}
\adir{\psi}{\hat{A}}=Tr\left(\hat{A}\dir{\psi}\otimes\dirc{\psi}\right)=Tr\left(\hat{A}\hat{\rho}\right)
\label{eq4}
\end{equation}

Trace in this equation can be represented as a sum or an integral according to cardinal number of representation basis. 
\[
Tr\left(\hat{A}\hat{\rho}\right)=\sum_{n,m}{A_{n,m}\rho_{m,n}} = \int{A\left(x',x)\right)\rho\left(x,x')\right)dxdx'}
\]

  Observables characterizing studied object classically, are usually represented by a function $A\left(x,p\right)$ depending on coordinate and momentum. Such function has a Fourier transform $a\left(P,Q\right)$
\[
 \begin{array}{l}
	 A\left(x,p\right)= \int{a\left(P,Q\right) \exp{\left(iPx-iQp\right)}\frac{dPdQ}{2\pi}};
\\ 	 a\left(P,Q\right)=
	 \int{A\left(x,p\right) \exp{\left(-iPx+iQp\right)}\frac{dxdp}{2\pi}}.
  \end{array}
\]
Operator $\hat{A}$ being appropriated to classical observable, can be generated by Weil correspondence rule through operator Fourier invertion
\[\hat{A}=
 	 \int{a\left(P,Q\right)\exp{\left(i\frac{P\hat{x}-iQ\hat{p}}{\hbar}\right)}\frac{dPdQ}{2\pi \hbar}}.
\]
  Taking into account the physical interpretation of transformation parameters $P$ and $Q$, one must include the Planck constant to its definition
\begin{equation}
	\hat{A}=
 	 \int{A\left(x,p\right) \exp{\left(i\frac{P\left(\hat{x}-x\right)-Q\left(\hat{p}-p\right)}{\hbar}\right)}\frac{dPdQ}{2\pi \hbar}\frac{dxdp}{2\pi \hbar}}
\end{equation}
This definition of quantum observabe operator produces the rule for  calculation of average values through classical function of observable
\begin{equation}
Tr\left(\hat{A}\hat{\rho}\right)  
=\int{A\left(x,p\right) Tr\left[\exp{\left(i\frac{P\left(\hat{x}-x\right)-Q\left(\hat{p}-p\right)}{\hbar}\right)}\cdot\hat{\rho}\right] \frac{dPdQ}{2\pi \hbar}\frac{dxdp}{2\pi \hbar}} 
\label{eq1}
\end{equation}

 One can define the kernel of integral (\ref{eq1}) as the special form of density operator, named the Wigner function
\begin{equation}
W\left(x,p\right)  
=\int{Tr\left[\exp{\left(i\frac{P\left(\hat{x}-x\right)-Q\left(\hat{p}-p\right)}{\hbar}\right)}\cdot\hat{\rho}\right] \frac{dPdQ}{\left(2\pi\hbar\right)^2 }}. 
\label{intdef}
\end{equation}
This definition makes it possible to represent the average value in the classical-like form
\begin{equation}
Tr\left(\hat{A}\hat{\rho}\right) =\int{A\left(x,p\right)W\left(x,p\right)dxdp}.
\label{eq3}
\end{equation}
Classical expression for average value 
\begin{equation}
\overline{A\left(x,p\right)} =\int{A\left(x,p\right)\rho\left(x,p\right)dxdp},
\label{eq2}
\end{equation}
contains the probability distribution function $\rho\left(x,p\right)$. 

Analogy between (\ref{eq3}) and (\ref{eq2}) can be the argument for interpretation of Wigner function as quantum probability distribution function, but Wigner function often has negative values that are inadequate to probability interpretation.

Typical examples that demonstrate the negativity of Wigner function are excited states of quantum oscillator and Schredinger Cat states.

The Wigner function of the first excited state of the oscillator is
\[
	 W_1\left(x,p\right)
	=\frac{1}{2\pi \hbar}\left(\frac{x^2}{\sigma_x^2}+ \frac{p^2}{\sigma_p^2} -1\right)e^{\left(-\frac{x^2}{2\sigma_x^2}- \frac{p^2}{2\sigma_p^2} \right)}
\]
Here $\sigma_x$ and $\sigma_p$ are uncertainties of coordinate and momentum. They satisfy the 
uncertainty relation $\sigma_x\sigma_p=\frac{\hbar}{2}$.

The above Wigner function is negative inside the domain $\frac{x^2}{\sigma_x^2}+ \frac{p^2}{\sigma_p^2} <1$.

\begin{figure*}[h!]
 \begin{center}
	\epsfbox{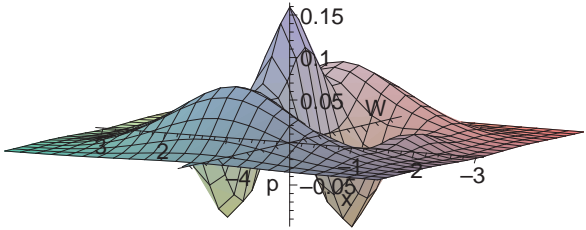}
	\caption{The Wigner function for the Schredinger Cat with distance between components $a=3\sigma_x$.}
	\label{cat}
\end{center}
\end{figure*}

The Schredinger Cat state with coherent components 
\[
	\psi_0\left(x\right)=\left(\frac{1}{2\pi \sigma^2}\right)^{1/4}\exp\left(-\frac{x^2}{4\sigma^2}\right)
\]
is
\[
	\psi_{cat}\left(x;a\right)=\frac{\psi_0\left(x-a\right)+\psi_0\left(x+a\right)} {\sqrt{2\left( 1+\exp{\left(-\frac{a^2}{2\sigma^2}\right)}\right)}}
\]
Its Wigner function is showen in figure \ref{cat}. One can see that it has domains of negativity.

These examples demonstrate clearly that Wigner function can not be interpreted as density of probability distribution. It can not specify probability for an arbitrary domain of phase space through standard equation $dP=W\left(x,p\right)dxdp$ because of the possible negativity. 

Additional insight  to the properties of the Wigner function follows from the  
 expression (\ref{eq1}) by displacement operator
\[\hat{D}\left(P,Q\right)=\exp{\left(i\frac{P\hat{x}-Q\hat{p}}{\hbar}\right)}
\]
 Average of displacement operator, named the Weyl function  
\[ \widetilde{W}\left(P,Q\right) =\frac{1}{2\pi \hbar}\mdir{\hat{D}\left(P,Q\right)}
=\frac{1}{2\pi \hbar} Tr\left[\exp{\left(i\frac{P\hat{x}-Q\hat{p}}{\hbar}\right)}\cdot\hat{\rho}\right], 
\]
 is the Fourier transform of Wigner function. As the Fourier inversion of the Weyl function
\begin{equation}
W\left(x,p\right)=\int{\mdir{\hat{D}\left(P,Q\right)}
\exp{\left(-i\frac{Px-Qp}{\hbar}\right)}
 \frac{dPdQ}{2\pi \hbar}},
\label{eqD}
\end{equation}
the Wigner function must have special properties. For instance, it can not take zero values on any domain and so in the case of zero probability for particle registration in any domain the Wigner function can not reflect the probability density. 

At the same time calculation of averages through (\ref{eq3}) makes evidence of ability of possibility to apply the Wigner function as density of probability distribution. To resolve this contradiction, we have to reassess the definition of probability for quantum phenomena.

\section{Probability Fundamentals}

The probability theory is the mathematical model describing properties of the sets of repeating measurable values. It includes the features of these sets and methods for analyzing but does not consider the possible causes of the spread of measurable values.  Common opinion is that spread of values results from inaccuracy of measurements and the increase of accuracy causes the decrease of spread. This assumption is one of the underlying principles of classical physics.

Quantum phenomena are distinguished by the existence of lower limit of the spread. This limit is expressed by form of uncertainty relations and its existence  is in contradiction with classical assumptions. Thus distinguishing feature of quantum physics is the existence of special, quantum properties of the sets of repeated measurable values. Since features of those sets are prerequisites of the theory of probability they can not be deduced from this theory are to be used at formation of the base of the probability theory for quantum phenomena. 

Main question is ine how to define the density of probability distribution for quantum phenomena on the phase space. The probability theory provides the definition for the probability density for those distributions only where density exists. Possibility of the probability density to exist or not to exist follows from applicability of two limit transitions in the theoretical basis of probability theory.

Theory of probability is based on a space of events which is the composition of set of events of topology base. The probability is defined as an amount giving a prognosis for potential realizability of events in each subset from base. This primitive explanation makes need in a strong formulation guiding the measurement analysis.

The first limit transition is used at constructing the probability definition as a 
limiting value of the frequency of events at non-limited increase of the 
volume of the sample for a finite event space. 

 When the space of events has a finite basis $U_k:\  U=\bigcup{U_k}$, numerical enough tests result in repetitons of each event. Number of repetitions $N_k$ actually depends on the number of tests  $N$.
 Relative frequency $\nu_k=\frac{N_k}{N}$ has a weak dependence vanishing for infinite test number. The limit of the relative frequency is probability.
 
  In real measurements number of tests is always finite. Consequently each real measurement  gives not probability but its estimated value only. The  value of probability got through measurement is the supposition:
\[
	N\rightarrow\infty\Rightarrow \forall_{k=1\ldots L}\  \exists\  P_k = \lim{\nu_k}
\]

 When $P_k=0$, $N_k < 0$ is impossible $N_n \geq 0$ is possible. In other words, zeroth probability does not lead to zeroth  number of events, it tolerates any finite non-negative number.
 
 Necessity of such a transition results in the fact that the idea of the probability itself can 
not be specific for real world, it is specific for theoretical models only. 
In real measurements number of tests is always finite, and the properties of 
the relative frequency for a given event in an infinite number of 
measurements remain to be always not more than supposition. Real value of relative frequency of each event can differ from its probability. The difference $\nu_k - P_k$ is often considered as unremovable error of measurement.

Physical meaning of  this fact is that the theoretical describtion of the properties of a phenomenon has a uncertainty. This uncertainty is the fundamental property of random phenomena and can not be eliminated.

The second limit transition is used in the definition for the probability 
conception on a continuous event space. This definition is implemented in 
two steps. At the first one the space of events $U$ is endowed a topology 
base with a finite number of subsets $U_k:\  U=\bigcup{U_k}$.  Relative frequency $\nu_k$ is defined through number of repetition $N_{x\in U_k}$ throughout the entire subset $U_k$ and probability is defined for each subset completely
\[
	 U=\lim_{L\rightarrow\infty}{\bigcup^L_{k=1}{U_k}} \Rightarrow P_k=\lim_{N\rightarrow\infty}{\frac{N_{x\in U_k}}{N}}.
\]
\begin{figure*}[h!]
\begin{picture}(400,55)
\put(2,2){\line(1,0){140}\line(0,1){50}}
\put(72,2){\line(0,1){50}}
\put(2,52){\line(0,-1){50}\line(1,0){140}}

\put(202,2){\line(1,0){140}\line(0,1){50}}
\put(272,2){\line(0,1){50}}
\put(202,52){\line(0,-1){50}\line(1,0){140}}
\put(202,27){\line(1,0){140}}

\put(30,20){\large $U_1$}
\put(100,20){\large $U_2$}
\put(160,20){\huge $\Rightarrow$}
\put(210,30){ $U_{1.0}\rightarrow U_1$}
\put(210,10){ $U_{1.1}\rightarrow U_3$}
\put(280,30){ $U_{2.0}\rightarrow U_2$}
\put(280,10){ $U_{2.1}\rightarrow U_4$}
\end{picture}
\caption{ Parcelling of topology base}
\label{f_1}
\end{figure*}
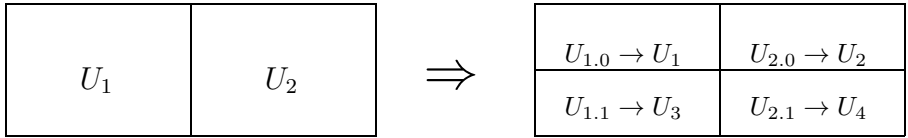

 At the second step the subsets of the topology base are 
infinitely grained as is showen in figure \ref{f_1}. 
\[
		U_k\rightarrow U_{k.0}\bigcup U_{k.1};\ 
		\left\{k.0,k.1;k=1\ldots L\right\}\Rightarrow\left\{k;k=1\ldots 2L\right\}.
\]
This makes it possible to define the probability density 
for subsets with small enough measure $\mu\left(U_n\right)$ as the limit of the ratio of 
probability and measure. 
\[
	p\left(x\right)=\lim_{\mu\left(U_k\right)\rightarrow 0}\frac{P_k}{\mu\left(U_k\right)}; \ x\in U_k
\]
This limit transition is similar to the method for definition of continuous 
distributions of physical characteristics of continuous medium. The value 
like mass density can be defined only till each ``physically small volume'' 
includes a large enough number of molecules.

 If the infinite graining of event space is physically meaningful, one can define the probability density, otherwise it is possible to define the probability distribution on the finite mesure only. 
 
 Consequently the definition of probability distribution is derived from the physically important properties of the event space. 
 
 Classical physics supposes the existence of material point - an object having negligible size. Such object is described as a point in phase space and this space has the Hausdorff topology.

Quantum physics is enforced to account of extention of each object through the uncertainty priciple. This priciple is mismatched with the Hausdorff topology of phase space and claimes for the reassessment of quantum phase space topology. 

\section{Measurement: simple model detector}

\begin{figure*}[h!]
	\begin{center}
	\epsfbox{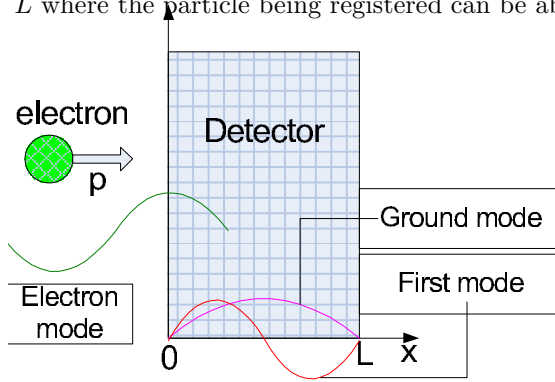}
	\caption{Model detector for the coordinate and the momentum measurement}
	\label{detector}
\end{center}
\end{figure*}

Origin of quantum phase space topology may be established through measurement interpretation. It needs a simple model for the joint measurement of noncommuting observables, such as coordinate and momentum. The simplest model is the registration of a particle by a showen in figure \ref{detector} model detector
 formed by a plate of given thickness $L$ where the particle being registered can be absorbed. 

Uncertainty of such coordinate measurement, characterized by standard deviation is equal to $\sigma_x=\frac{L}{2\sqrt{3}}$. After being absorbed, the particle goes to one of bound states pertinent to the system of states of rectangular potential well and has a momentum $p_k=\frac{2\pi \hbar}{L} k$.

It is obvious that the maximally exact measurement of momentum corresponds to such a design of the device that the separate detected event is registration of the number $k$ of each specific mode. The uncertainty of such registration is the standard deviation for discrete readings, this is $\sigma_k=\frac{1}{2\sqrt{3}}$. For the momentum one has, respectively, the estimation of the typical uncertainty $\sigma_p=\frac{\pi \hbar}{\sqrt{3}L}$.

The product of uncertainties is $\sigma_p \sigma_x=\frac{\pi}{3}\frac{\hbar}{2}$.
This estimation is almost equal to the limit $\frac{\hbar}{2}$ given by the uncertainty principle and experimentally separable events can not have the phase space domains smaller than $\frac{\hbar}{2}$.
  
Thus topology base of phase space must be generated by domains $U_k$ having the measure not smaller than $\frac{\hbar}{2}$ only. 

Respective propability distribution can be defined for the whole domain only and is undefined for any part of domain. 

\section{ Measurement of probability distribution on phase space}
Definition of probability distribution on phase space is constructed through mapping of space of measured events to the phase space. In the case of the above detector mapping function has a fixed value inside of the domain $0\leq x \leq L$, which represents the detector plate, and zero value outside it. Each of inner states of particle is distinguished by momentum and mapped onto domain around the  momentum of state $p_k=\frac{\pi \hbar}{2L}k$. Upon supposition that all states are distinguishable and set of states is full, one has to supposed that each state is mapped by the rectangular domain 
\[U_k=\left[0\leq x \leq L\right] \cap \left[\frac{\pi \hbar}{2L}\left(k-\frac{1}{2}\right) \leq p \leq \frac{\pi \hbar}{2L}\left(k+\frac{1}{2}\right)\right].\]
Overlap of the boudaries  is irrelevant because of null measure.

Classical measurement  for each domain can be realized as measurement of the special 
independent observables - observable of belonging of the coordinate to the interval $\left[0,L\right]$ and observable of belonging of the momentum to the interval $\left[p_k-\frac{1}{2}\Delta p,p_k+\frac{1}{2}\Delta p\right]$
\[
\begin{array}{l}
	X\left(x\right)=\frac{1}{L}\left(\theta\left(x\right)-\theta\left(x-L\right)\right)\\
	\Pi_k\left(p\right)=\frac{2L}{\hbar}
	 \left(\theta\left(p-p_k+\frac{1}{2}\Delta p\right) -\theta\left(p-p_k-\frac{1}{2}\Delta p\right)\right)
\end{array}
\]
Quantum observables because of uncertainty principle can not be independent. In the case of joint measurement quantum version of second observable forces the replacement of the arbitrary value $\Delta p$ by the quantum-conditioned interval  $\frac{\pi \hbar}{2L}$.  
\[
	\Pi_k\left(p\right)=\frac{2L}{\hbar}
	 \left(\theta\left(p-\frac{\pi \hbar}{2L}\left(k-\frac{1}{2}\right)\right) -\theta\left(p-\frac{\pi \hbar}{2L}\left(k+\frac{1}{2}\right)\right)\right)
\]
Thus it is needed to involve to consideration the observable for belonging to such domains $U_k$ of phase space that can  have size not smaller than $\frac{\hbar}{2}$.  
It is defined by means of Weil rule as
\[
	\hat{U}_k=
 	 \int{X\left(x\right)\Pi_k\left(p\right) \exp{\left(i\frac{P\left(\hat{x}-x\right)-Q\left(\hat{p}-p\right)}{\hbar}\right)}\frac{dPdQ}{2\pi \hbar}\frac{dxdp}{2\pi \hbar}}.
\]
As a result value of propability for domain $U_k$ can be defined as
\begin{equation}
	P_k=\mathop{\int\int}_{x,p \in U_k}{W\left(x,p\right)dpdx}
\label{defP}
\end{equation}
while each domain has the size not smaller than $\frac{\hbar}{2}$ only. 

So, probability distribution on phase space is meaningful till the topology base consists of the domains with the bounded below size only.

The main concept we get is that (\ref{defP}) has the sense of probability till it remains non-negative only.

The different interpretation is that Wigner function gets the sense of probability after the reduction to bounded-below domains only. 

\section{Gaussian Smoothing and Husimi function}

Disadvantage of reduction of the phase space topology to rectangular domains is in the use of the basis of the wave functions with a compact carrier. It is well known that the use of such basis for relativistic quantum field leads to emergence of irreducible representations and  to loss of uniqueness of vacuum state.

The stronger method of reduction consists of the smoothing of Wigner function with  non-vanishing weight function such as Gaussian function.
Generally, it can be done by the Gaussian distribution with arbitrary halfwidthes both of coordinate $\sigma_x$ and momentum $\sigma_p$. 
\[W_{smooth}\left(x,p;\sigma_x,\sigma_p\right)=\int{
W\left(x',p'\right) \exp{\left(-\frac{\left(x'-x\right)^2}{\sigma_x^2}-\frac{\left(p'-p\right)^2}{\sigma_p^2}\right)}
\frac{dx'dp'}{2\pi\sigma_x\sigma_p}}
\]
Product of halfwidthes $\sigma_d=\sigma_x\sigma_p$ defines the effective measure of smoothing.
In the case this measure corresponds to the uncertaity principle $\sigma_d =\frac{\hbar}{2}$ the Gaussian smoothing results in the Husimi function
\[W_{up}\left(x,p\right)=Q\left(\alpha\right);\ \alpha=\frac{x}{\sigma_x}+i\frac{p}{\sigma_p}
\]
while in general case this representation can not take place. 

Since Hisimi function is determined as the average of the density operator by coherent state
\[Q\left(\alpha\right)=\adir{\alpha}{\hat{\rho}},
\] 
it remains non-negative for density operator of arbitrary quantum state. The main corollary of this fact is that the Gaussian smoothing with quantum measure $\sigma_d =\frac{\hbar}{2}$ restores probability interpretation of Wigner function.

\section{ Criteria of Nonclassicality}

Nonclassicality is topical property for quantum objects since this peculiarity be applied for realization of various essentially quantum devices. It can be declaratively understood as being opposite to quasiclassicality. Practical applications of nonclassicality require the ovservable values specific for it. 

As a peculiarity of quantum state nonclassicality is a property of density operator and must have a suitable expression through some representation of this.

 One of special nonclassical properties is the negativity of Wigner function. This has place when the probability of detection turns to zero for some domain in phase space. Since Wigner function has the properties of the Fourier transformation it can not vanish on any finite domain and must have negative values within the domain of the vanishing probability. So negativity of Wigner function is the criterion of nonclassicality. Weakness of this criterion is in its unpracticality. It requires the full information about the properties of the density operator and can require numerous measurements.

In the broad interpretation nonclassicality is opposite to the quasiclassicality and can be described by difference between given quantum state and some quasiclassical one.
Quasiclassical states are eigenstates of the decreasing operator 
\[\hat{a}=\frac{\sigma}{\hbar}\hat{x}+i\frac{1}{2\sigma}\hat{p}
\]
Main property of quasiclassical state is that it has the good defined average of this operator
\[\hat{a}\dir{\alpha}=\alpha\dir{\alpha} \Rightarrow \adir{\alpha}{\hat{a}}=\alpha
\]
and can describe states with vanishing uncertainties both of coordinate and momentum. 

Next significant property of quasiclassical state is that it has a good defined deviation
\[\sigma_p^2=\adir{\alpha}{\hat{p}^2}-\adir{\alpha}{\hat{p}}^2=\sigma^2;\ 
\sigma_x^2=\adir{\alpha}{\hat{x}^2}-\adir{\alpha}{\hat{x}}^2=\frac{\hbar^2}{4\sigma^2},
\]
and can characterized by elliptic domain of phase space with semi-axises $\sigma$, $\frac{\hbar}{2\sigma}$ and measure $\frac{\hbar}{2}$.

Difference between the given and some quasiclassical state depends on the parameters of the last one.
It is reasonable that average values for quasiclassical state must be equal to those of the given one.
Choise of deviation for the quasiclassical state is less trivial. In the case a given state has both coordinate and momentum deviations producting a minimum of uncertainty product $\sigma_x \sigma_p=\frac{\hbar}{2}$ such state can be quasiclassical only. So nonclassical state must have the uncertainty product strictly greater than the minimum one. Hence it is possible that quasiclassical state with some value of parameter $\sigma$ gives the best approximation of the given one. 

The rating of the approximation accuracy followes from the expansion of given state in series 
\[\dir{\psi}=\sum_{n=0}{\psi_n\left(\alpha,\sigma\right)\dir{\alpha,\sigma;n}}
\]
on shifted coherent basis $\dir{\alpha,\sigma;n}=\frac{\left(\hat{a}-\alpha\right)^n}{\sqrt{n!}}\dir{\alpha}
$.
Since this basis is orthogonal, difference between given and quasiclassical states has the norm 
\[\left\|\dir{\psi}-\dir{\alpha}\right\|^2=1-\left|\psi_0\left(\alpha,\sigma\right)\right|^2.
\]
When this norm is equal to zero, given state is completely quasiclassical. 

The part of nonclassical components in series can be estimated by the average of the exitation number $\hat{n}$ 
\[
\overline{n}=\adir{\psi}{\hat{a}^+ \hat{a}}=\sum_{n=1}{n\left|\psi_n\left(\alpha,\sigma\right)\right|^2}.
\]
This average can be expressed by the deviations for the given state  
$\overline{n}= \frac{\sigma_p^2}{\sigma^2}+\frac{\sigma^2\sigma_x^2}{4\hbar^2}-\frac{1}{2}$.
Best approximation for given state corresponds to the case of minimum of this expression and is
$\overline{n}=\frac{\sigma_x\sigma_p}{\hbar}-\frac{1}{2}$.
This value can work as the criterion of nonclassicality since it turns to zero for classical states only. 

Physical meaning of thist criterion is the excess size of phase space domain occupied by given state over the minimum size occupied by quasiclassical one. 

\section{Conclusion}

The Wigner function as a special representation of density operator retains the probability interpretation on the space of its own definition. This is the quantum phase space differing from the classical one by the topology base. Peculiar to classical phase space Hausdorff topology changes to bounded-below base in the quantum case.

The probability density distribution retains the meaning on the quantum phase space through smoothing only. This smoothing must be realized on domains with size not smaller than quantum minimum $\frac{\hbar}{2}$ .

Negativity of the Wigner function is the phenomenon of zero probability for some domains of phase space containing negative values of Wigner function. 

The effective observable of nonclassicality of given state is the size of phase space domain occupied by it. The state remains quasiclassical till the occupied domain has the minimum size.  
Exceeding of that size results in negativity of the Wigner function  and in other phenomena of nonclassicality.


\end{document}